# Multi-extremum optimization in lens design: navigation through merit function valleys maze.

Ilya P. Agurok

Department of Electrical and Computer Engineering, UC San Diego, 9500 Gilman Drive, La Jolla, CA 92093.
Corresponding author: iagurok@ucsd.edu

**Abstract.**

Lens designers routinely use optimization in their everyday practice. Local optimization algorithms lead to the nearest minimum. For comprehensive research on lens architecture, ZEMAX offers two options for multi-extremum optimization: Global and Hammer. They provide a number of solutions depending on the designer's choice for starting point. Both Global and Hammer optimization options are stochastic in nature and cannot ensure completeness of the result. In this paper, a new deterministic approach for multi-extremum optimization is proposed. Optimal solutions for even moderate complexity optical architectures are shown to be located within extended merit function valleys. Merit function minimums are separated by saddle points. An effective algorithm to travel over these valleys from one local minimum through a saddle point to another minimum is proposed. From this new minimum, a new valley is found which leads through another saddle point to another minimum and so on. In a finite number of steps, a complete mutually connected system of stationary points (minimums and saddle points) are revealed, giving a reasonable assurance that the search is completed.

**Introduction.**

The design space of optical systems is a complicated multidimensional space, comprising of a number of optimal solutions (local minimums of the assigned merit function). In early work [1] 10 such local minimums were found using expert system based optimization even for a simple Cooke triplet. This multi-extremum optimization problem attracted close attention from the beginning of the computer aided lens design era. A number of effective algorithms were proposed, the majority of them stochastic. While they are able to solve the main practical problem, revealing a number of minimums, they cannot ensure that all minimums were found. An effective blow-up/settle-down (BUSD) algorithm was proposed in Ref. [2]. At the first, a local minimum is found depending on the user's choice of the starting point. After that, BUSD forces the design to "blow-up", thereby changing the values of the optimization parameters significantly. This is sufficient to escape the 'gravity' of the already known local minimum and the following local optimization will "settle-down" the search to a new one. The direction of "blow-up" step is the direction of Dumped Least Square (DLS) [3] method searching for the maximum. In Ref. [4], Optical Research Associates announced a global optimization option for their Code V lens design software but did not give any details on its operation principals. The described behavior is similar to that which was shown in Ref. [2]. Simulated annealing is another stochastic algorithm, which uses random steps at every cycle and accepts all steps which result in a reduction of optimization criterion and others with some probability. This tactic prevents the search algorithm from losing solutions in the areas separated from the starting point with large values of the merit function [5]. ZEMAX commercial software has two multi-extremum optimization options: Global and Hammer optimization. Global optimization uses genetic algorithm techniques with fast local optimization solutions upgrade [6]. Hammer algorithm uses heuristic parameters adjustment and optimization. However both options are stochastic and without a guarantee that the deepest minimum was found.

In Ref. [7], an escape function to the global optimization method was proposed. In this method, the first minimum is found with the use of local optimization. This minimum creates a crater in the multidimensional optimization space. The special escape function has two adjustable parameters and is added to the optimized function in order to fill up the minimum crater, thus eliminating any already found minimum from consideration. The next local optimization will lead to the next minimum. The problem is finding appropriate values of these two parameters to fill the crater smoothly, without creating a new artificial minimum. Authors found several solutions to this problem, and in a design example found 50 solutions for their six-elements lens. Nevertheless, the exact universal solution for the escape function parameters was not found in this paper nor in later developments [8]. It is the first determinative algorithm in the row of proposed global optimization strategies. While the initial starting point can be chosen arbitrary, the algorithm was designed to sequentially find all minimums.

In the recent years, promising results were reported in the systematic search of minimums for multi-extremum optimization of optical systems by consequent search of a new minimum through the closest saddle points [9]. Based on the general topology consideration in Ref. [9], saddle points were shown to be points of transition between neighboring minimums. Transitioning between minimums includes two general steps: saddle point detection (optimization from the minimum to saddle point) and local optimization to a new minimum. In the Ref. [9] such a saddle point detection (SPD) algorithm was proposed. Local optimization to a new minimum was not discussed in detail, thus was probably conventional. Such methodology gives an opportunity to reveal new minimums in a sequential and systematic way. A closed system of minimums mutually bounded by saddle points has a high probability that the system is complete. Cooke triplet global optimization was reported using this method. Moreover, in Ref. [10, 11] it is shown that with addition to the optical system a neutral null-element it creates a new saddle point and can pave the way to new minimums.

However some specific aspects related to optimization of more complex systems are not discussed. It is known that optimization of optical systems has a specific problem, an ill-conditioning of the Hessian matrix [3, 12, 13]. Because the Hessian matrix is ill-conditioned, deep multidimensional valleys on the optimization field exist. Local optimization methods do not work properly in this case. To solve this problem the method of conjugated gradients was proposed [12] but only works up to a certain point of valley complexity and then can fail. Newton methods used dumped least square algorithm (DLS) [3, 13] which suppresses dependable parameters but will come straight to the closest point of the valley bottom instead of the minimum. Both methods give some relief to the stated problem, but a more deep consideration must be taken. This paper demonstrates that a very specific method of traveling through the bottom of the valleys has to be developed to reveal a structure of the merit function. This method leads to a minimum rather than to the closest merit function valley bottom. It then reveals saddle points and leads to a new minimum sequentially paving a way to achieving a systematic multi-extremum search.

## 1. Objective for optimization. Signature features of the merit function landscape.

As an example, a ZEBASEV K_002 $20^x$ microscope objective was chosen for optimization. Its layout is shown in Fig. 1.

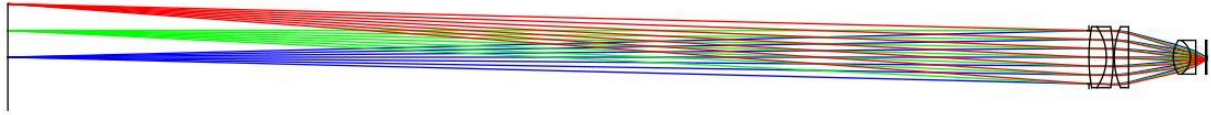

a) Objective layout.

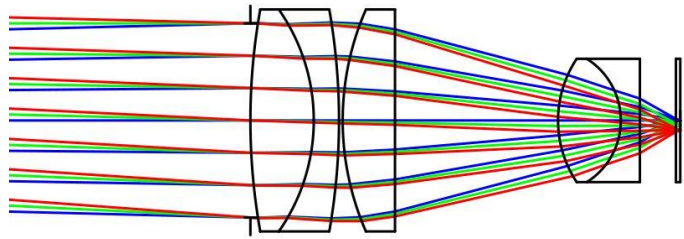

b) Zoomed view.
**Fig. 1 Microscope objective K_002 from ZEBASEV.**

The optical prescription is shown in the Table 1. All glasses are of Schott preferred type from the 2019 catalog. The eight optimization parameters are marked with the symbol "v" - six radii and two air gaps. In the proceeding text, radii will be denoted as Ri and air gaps as Ti. In order to keep the lens manufacturable, avoid thin edges, and negative air gaps and other problems, optimization parameters have the following constraints.

$$\begin{aligned}
&\text{abs}(R2) >= 10.0 \text{ mm}, \\
&\text{abs}(R3) >= 7.7 \text{mm}, \\
&\text{abs}(R4) >= 11.0 \text{ mm}, \\
&\text{abs}(R5) >= 8.0 \text{mm}, \\
&\text{abs}(R7) >= 4.0 \text{ mm}, \\
&\text{abs}(R8) >= 3.0 \text{ mm}, \\
&4.0 \text{ mm} < T6 < 8.2 \text{ mm}, \\
&T9 >= 0.3 \text{ mm}.
\end{aligned} \qquad (1)$$

The aperture stop (entrance pupil) is located at surface 1. The Entrance pupil diameter is 8 mm and the objective operates with F# = 1.02 at the image space. Spot diagrams at three image heights: -0.4 mm (object height 8 mm), -0.2 mm (object height 4 mm) and 0 mm are shown in Fig. 2. The size of spots are shown in microns. The size of the square field in Fig. 2 is 20 um while all the remaining spot diagrams in the paper are 10 um. To reduce computational burden associated with raytracing derivatives test optimization was made at a single wavelength of 0.587 um.

**Table 1 Optical prescription of the K002 objective.**

| Surface | Radius | Thickness | Material | Semi-Diam. |
|---|---|---|---|---|
| OBJ | Infinity | 162.8140 | | |
| STO | Infinity | 0.0000 | | 4.0 |
| 2 | 25.8691 V | 2.6182 | K10 | |
| 3 | -7.9612 V | 1.0414 | SF1 | |
| 4 | -26.2169 V | 0.1524 | | |
| 5 | 11.3792 V | 2.1590 | N-SK5 | |
| 6 | Infinity | 6.7313 V | | |
| 7 | 4.5770 V | 2.5908 | N-SK5 | |
| 8 | -3.0002 V | 0.7874 | F5 | |
| 9 | Infinity | 1.4882 V | | |
| 10 | Infinity | 0.1778 | N-K5 | |
| 11 | Infinity | 0.0000 | | |
| IMA | Infinity | | | |

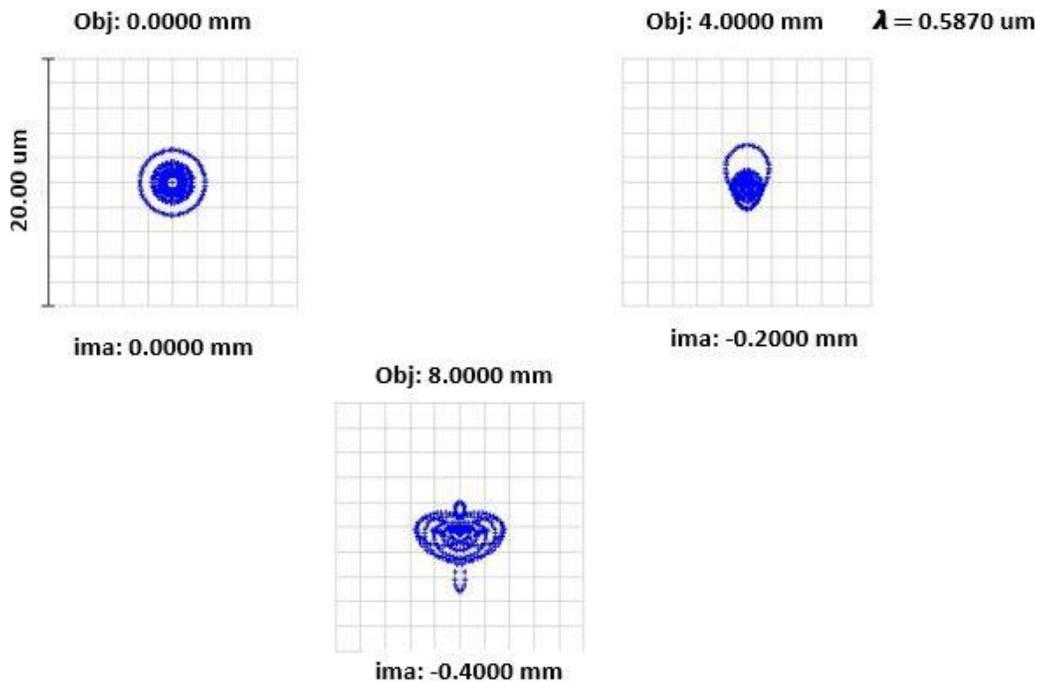

**Fig. 2 Spot diagrams.**

To estimate the image quality, each beam consists of 79 rays (5 rays at the pupil radius) for each field point were raytraced. The criterion of optimization, C, was a sum of squared lateral aberrations plus constraints violation penalty function.

$$C=\sum_{m=1}^{3}\left\{\sum_{i=1}^{79}[(xi-x00m)^2 + (yi-y00m)^2]\right\} + P; \qquad (2)$$

where m is beam number, xi -x ray coordinates at the image plane, yi- y coordinates, x00m – paraxial x image coordinate (for all beams $x00_m$=0.0 mm) and y00m -paraxial y image coordinates at the image plane ( $y00_0$= 0.0 mm; $y00_1$= -0.2 mm and $y00_2$= -0.4 mm). So, criterion C depends on both lateral aberrations and current magnification. P is a penalty function which is zero inside constraints area and grows fast in the case of constraints violation.
For radii:

$$P= 0.0 \text{ inside constraints and}$$
$$P= 0.25*(RIConstraint-RI)^2 \text{ outside constraints} \quad (3a)$$

For airgaps:

$$P= 0.0 \text{ inside constraints and}$$
$$P= (TIConstraint-TI)^2 \text{ outside constraints} \quad (3b)$$

where RIConstraint- constraint radii and TIConstraint – constraint thicknesses. For the ZEBASE K_002 objective, C= 6.08E-4, indicating that the lens is well optimized. Attempts for further improvement with ZEMAX local optimizations (DSL or Orthogonal descent resulted in wobbling around this point without sensible criterion improvements. A quadratic model of C in the vicinity of the start point can help reveal the reason of such local optimization behavior. First and second derivatives of the criterion function were calculated using the finite difference method. The use of radii in the optimization makes it impossible to overstep the sign barrier. So, curvatures and air gaps will be used as optimization parameters. For derivative calculations, air gap increments were 3 um and increments for curvatures were 0.00002 $mm^{-1}$. The quadratic model of the criterion C is

$$C(x_i) = C_0 + g^T(x_i)*(\Delta x_i) + 0.5*(\Delta x_i^T)Q(\Delta x_i) ; \quad (4)$$

where $g^T$ is transposed vector of first derivatives, $\Delta x_i$ – vector of parameter increments, and Q is Hessian matrix of second derivatives. The Hessian matrix has a diagonal symmetry. For such matrices, linear algebra states that rotations of coordinate system make matrix Q diagonal or

$$C(u_i) = C_0 + \sum_{i=1}^{8} gui * \Delta ui + 0.5 * \sum_{i=1}^{8} Ei * (\Delta ui)^2 \quad (5)$$

where $\Delta u_i$ parameters increments in rotated coordinates system, $gu_i$ are first derivatives in rotated coordinate system, and $E_i$ eigen values of the Hessian matrix. For the optical prescription (Table 1) Eigen values are shown in the Table 2.

**Table 2 Eigen values.**

| Number | 1 | 2 | 3 | 4 | 5 | 6 | 7 | 8 |
|---|---|---|---|---|---|---|---|---|
| Eigen | 3.67E+5 | 3.00E+2 | 1.98E+1 | 1.90E+2 | 2.54E-1 | 1.57E-1 | 6.67E-4 | 5.72E-6 |

Derivatives of variables in the rotated coordinates system are shown in the Table 3.

**Table 3 Derivatives.**

| Number | 1 | 2 | 3 | 4 | 5 | 6 | 7 | 8 |
|---|---|---|---|---|---|---|---|---|
| Derivat. | -9.2E-2 | 3.64E-5 | -1.3E-3 | 2.9E-3 | -1.6E-4 | 5.4E-4 | 2.2E-4 | 1.4E-4 |

The two last eigen values are very small indicating that the Hessian matrix is ill-conditioned. Across the six variables in the rotated coordinate system, the criterion function will be a fast-growing narrow parabola (Eq. (5)) and along the last two variables, the criterion landscape is some kind of slow changing valley. Fig. 3 shows a criterion $C(R_2, R_3)$ contour map in the area of the optimization starting point of (Table 1) with this valley.

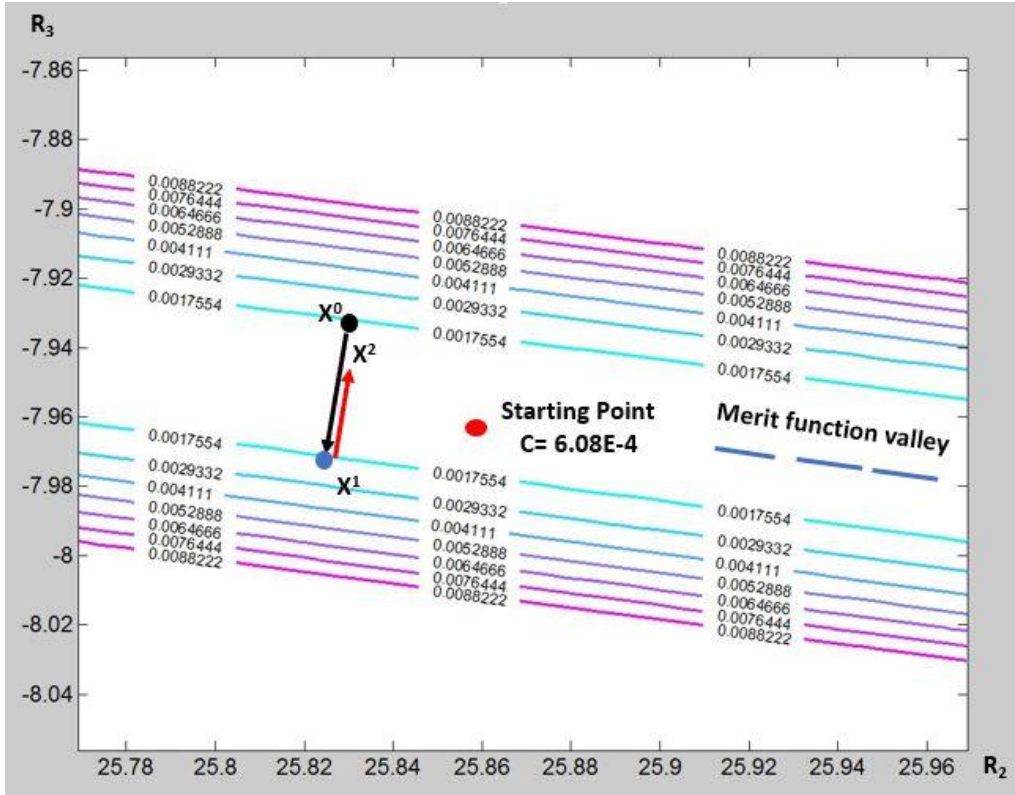

**Fig. 3 Contour map of the criterion C in the vicinity of the starting point.**

**II. Classical local optimization methods.**

There are two basic methods used in local optimization of nonlinear functions: gradient descent method [3, 12, 13] and Newton method [3, 13]. In our case, consequent gradient vectors ($X^0 X^1$ and $X^1 X^2$) will be counter collinear (Fig. 3). Gradient method begins to oscillate and stop at some point close to the valley bottom. The position of this point depends on the location of starting point $X^0$. Criterion of the Newtonian dumped least-square method (DLS) is

$$\{\sum_1^8 [gui + Ei * (\Delta ui)]\}^2 + \delta * \sum_1^8 (\Delta ui)^2 = \min, \qquad (6)$$

or in the other words the squared sum of criterion first derivatives plus the weighted Euclidean norm of the step will be a minimum. $\delta$ is damping constant. As in a gradient method optimization, steps will be repeated while recalculating the quadratic model of Eq. (5) until convergence. The solution at every step for each orthogonalized parameter is

$$\Delta u_i = (-gu_i * E_i)/(E_i{}^2 + \delta) \tag{7}$$

hence with any $E_i$ close to zero (ill conditioned Hessian matrix) solutions still exist. DLS is tending to converge toward the closest to starting point valley bottom point where the squared sum of criterion first derivatives is minimal.

**III. Proposed optimization strategy to operate on the perplexed merit function landscape.**

In the vicinity of each minimum (Fig. 4) it is encircled with equimagnitude surfaces (surfaces having the equal value of the optimization criterion). Equimagnitude surfaces bulge out of minimums. At some point S, with criterion value $C_S$, equimagnitude surfaces will meet each other.

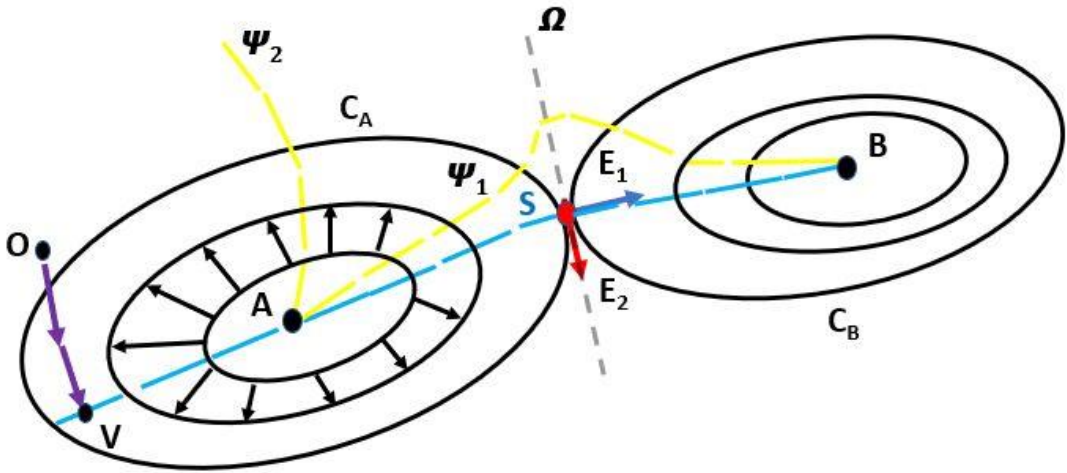

**Fig. 4** Multi-extremum search through the saddle point.

If we will go over the normal to the equimagnitude surfaces at the point S in both directions, we will enter equimagnitude surfaces encircling minimums (surfaces $C_A$ and $C_B$). The value of criterion will be less than $C_S$ for any small step in both sides. It is possible only if the gradient at the point S is zero and eigen value of the eigen vector $E_1$ parallel to the normal is negative. At the point S both equimagnitude surfaces have common tangential hyperplane $\Omega$. All points at this hyperplane are located outside equimagnitude surfaces $C_A$ and $C_B$ and have a criterion value lager than $C_S$. Hence, eigen values in this hyperplane are positive. So, at the point S first derivatives are zero, one eigen value negative while others are positive [9]. Such point is the saddle point of the Morse one type. Saddle points are separating areas of attraction to the neighboring minimums. The special roll of the saddle points in the stationary point networks was noticed at the first time in Ref. [14].

A gradient curve is a curve orthogonal to the equimagnitude surfaces at any point. Gradient curve ASB follows through the saddle point S. All other gradient curves connecting minimums A and B (for example curve $\Psi_1$) will inevitably step out of equimagnitude surfaces $C_A$ and $C_B$ and enter areas with criterion lager than $C_S$. So, the ASB curve has the lower maximum value of criterion besides the other gradient curves connecting minimums A and B. In other words, it is the path of slower growth leading from the minimum A toward saddle point and the path of slower descent toward minimum B. Merit (criterion) function valleys are those paths of slower growth/descent.

Moreover, there is no guarantee that any gradient curve originated from the minimum A rather than ASB will connect minimums. For example, curve $\Psi_2$ will not do so. Therefore, merit function valleys (ASB) are the only reliable path from the minimum A to the minimum B. In this paper starting from the initial point O, the local optimization will lead to some point at the closest valley bottom (point V in the Fig. 4). Then, the optimization will travel over the valley bottom until it will reach local minimum (point A). Further travel over the valley bottom leads to the saddle point. After that travel over the valley will lead to the next minimum. From this minimum, the optimization will pave the valley to a new saddle point and so on. By reaching constraints surface the closest minimum on the surface will be searched. If this minimum is separated from the previous one with the saddle point it will be an entrance to a new valley on the way back to the optimization space. This new valley will lead to a new minimum. Such tactics will be used in this paper. But in general, multi-extremum optimization has to be made on the constraint surface and all new valleys investigated.

Travel over the valley bottom will be performed in repeated cycles. Each cycle begins with DLS correction to the valley bottom. At the DLS step, the criterion $C_0$ will be calculated and using raytracing, finite differences technique vector g of derivatives and Hessian matrix Q of Eq. (4). Then using MATLAB eigen function will be calculated eigen vectors $V_i$ and eigen values $E_i$. The DLS step of Eq. (7) will be applied to the first six orthogonalized parameters and will not be applied to the last two dependable parameters at all. So the DLS step will lead to the closest point at the bottom of valley where derivatives across strong variables $u_i$ have to be zero (vertex of fast parabolas) but derivatives over weak variables with small eigen values can have some small value (slow growth or descent). The next operation will be a step over eigen vector providing the lower criterion increment. This lower criterion increment can be negative indicating descent to the minimum. Or it can be positive indicating travel over valley toward the saddle point. The optimization cycle will be repeated paving the path over the stationary points network.

In Ref. [9] the saddle point detection method (SPD) was proposed. In the SPD method, several arbitrary directions from the local minimum are chosen. In each direction a step will be made and then minimum of the criterion at the hyperplane orthogonal to the chosen direction will be found. After that a new step will follow from the minimum at the hyperplane and so on. Those minimums constitute the SPD curve. The maximum of the criterion over successful SPD curves will be the saddle point. The search in directions located within a wide enough solid angle will be successful. Searches in other directions may not. In this paper, saddle point optimization directions are better defined as directions of slowest growth (toward eigen vectors having a reasonably small eigen value).

### IV. Travel over R2 valley.

To travel over the valley, the leading variable will be chosen. It will prevent the optimization from wobbling. In this paper, the first leading variable will be a surface number two curvature and travel begins in the direction of increasing R2 curvature. All eigen vectors with negative projections on the first optimization parameter (R2 curvature) will be rotated 180º. Step in the curvatures/air gaps space was experimentally chosen as 0.003 (gaps were specified in mm's and curvatures in mm$^{-1}$ ). With travel over valley in the direction of increasing surface number two curvature criterion C at the beginning decreases until it reaches minimum with C=5.25E-4 and then increases until it will reach R3 (radius -7.7 mm) and T6 (4 mm) constraints surface with C=1.65E-3 without passing saddle points. The layout and spot diagrams at the R2 valley minimum are shown in the Fig. 5 and optical prescription in the Table 4.

**Table 4 R2 valley minimum.**

| Surface | Radius | Thickness | Material | Semi-Diam. |
|---|---|---|---|---|
| OBJ | Infinity | 162.8140 | | |
| STO | Infinity | 0.0000 | | 4.0 |
| 2 | 22.4880 | 2.6182 | K10 | |
| 3 | -8.1862 | 1.0414 | SF1 | |
| 4 | -28.3326 | 0.1524 | | |
| 5 | 11.5102 | 2.1590 | N-SK5 | |
| 6 | Infinity | 6.7163 | | |
| 7 | 4.4972 | 2.5908 | N-SK5 | |
| 8 | -2.9996 | 0.7874 | F5 | |
| 9 | Infinity | 1.3877 | | |
| 10 | Infinity | 0.1778 | N-K5 | |
| 11 | Infinity | 0.0000 | | |
| IMA | Infinity | | | |

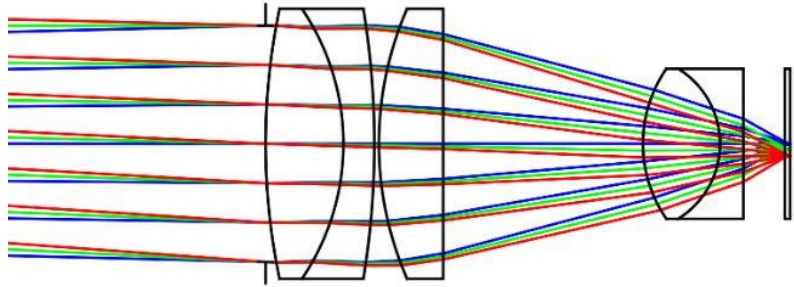

**a) R2 valley minimum layout.**

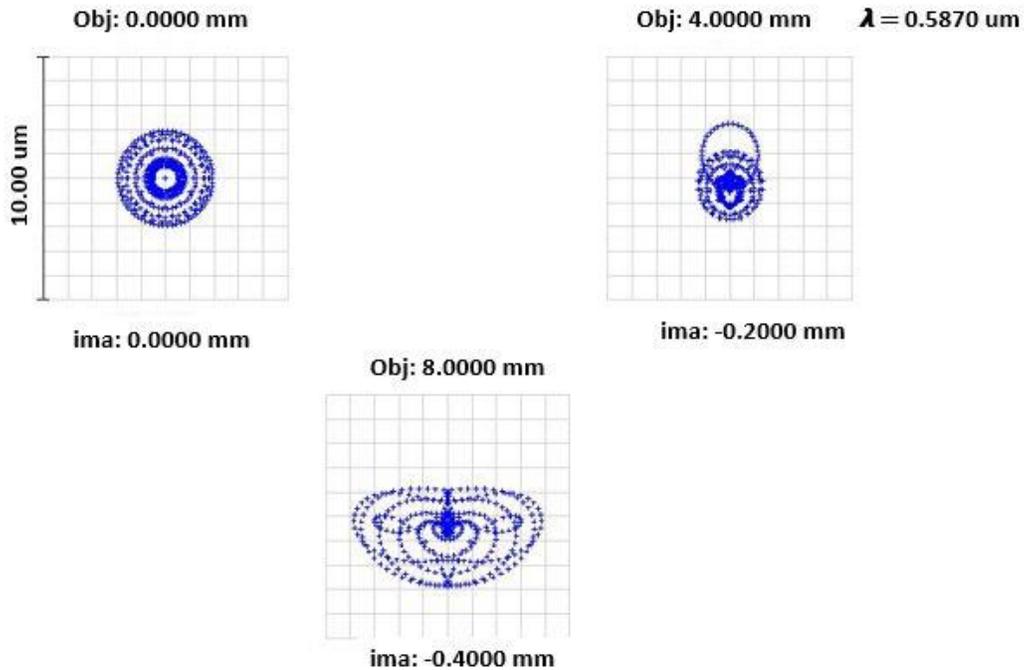

c) Spot diagrams.
Fig. 5 R2 valley minimum. C=5.25E-4.

To get some perception on the length of this valley let's look at the bottom points with slightly higher criterium C=5.3E-4. The difference in image quality (spot diagrams) between points having C=5.25E-4 and C=5.3E-4 look indistinguishable. Here is the point with lager R2 value which has C=5.3E-4.

**Table 5 Optical prescription at the point of the R2 valley with C=5.3E-4. Higher R2 value.**

| Surface | Radius | Thickness | Material | Semi-Diam. |
|---|---|---|---|---|
| OBJ | Infinity | 162.8140 | | |
| STO | Infinity | 0.0000 | | 4.0 |
| 2 | 23.9823 | 2.6182 | K10 | |
| 3 | -8.0991 | 1.0414 | SF1 | |
| 4 | -27.6506 | 0.1524 | | |
| 5 | 11.4687 | 2.1590 | N-SK5 | |
| 6 | Infinity | 6.7265 | | |
| 7 | 4.5293 | 2.5908 | N-SK5 | |
| 8 | -2.9998 | 0.7874 | F5 | |
| 9 | Infinity | 1.4559 | | |
| 10 | Infinity | 0.1778 | N-K5 | |
| 11 | Infinity | 0.0000 | | |
| IMA | Infinity | | | |

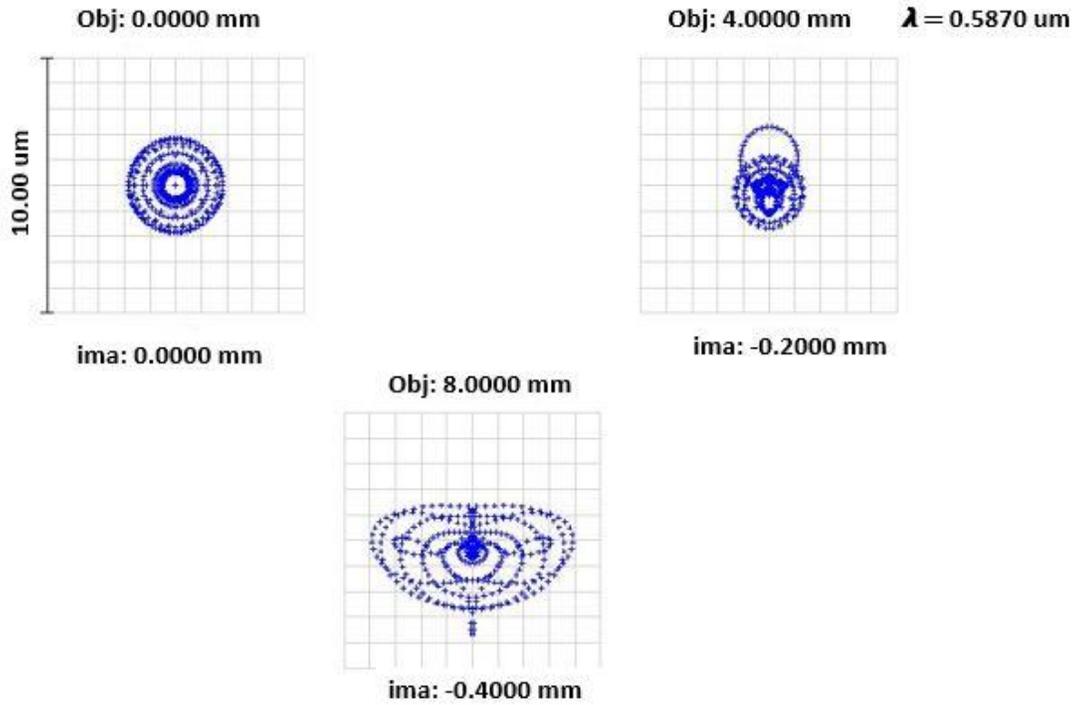

**Fig. 6 Spot diagrams.**

Here is the point with lower R2 value and again the same C= 5.3E-4.

**Table 6 Optical prescription at the point of the R2 valley with C=5.3E-4. Lower R2 value.**

| Surface | Radius | Thickness | Material | Semi-Diam. |
|---|---|---|---|---|
| OBJ | Infinity | 162.8140 | | |
| STO | Infinity | 0.0000 | | 4.0 |
| 2 | 21.3671 | 2.6182 | K10 | |
| 3 | -8.2355 | 1.0414 | SF1 | |
| 4 | -28.9572 | 0.1524 | | |
| 5 | 11.4951 | 2.1590 | N-SK5 | |
| 6 | Infinity | 6.6639 | | |
| 7 | 4.4990 | 2.5908 | N-SK5 | |
| 8 | -2.9996 | 0.7874 | F5 | |
| 9 | Infinity | 1.3461 | | |
| 10 | Infinity | 0.1778 | N-K5 | |
| 11 | Infinity | 0.0000 | | |
| IMA | Infinity | | | |

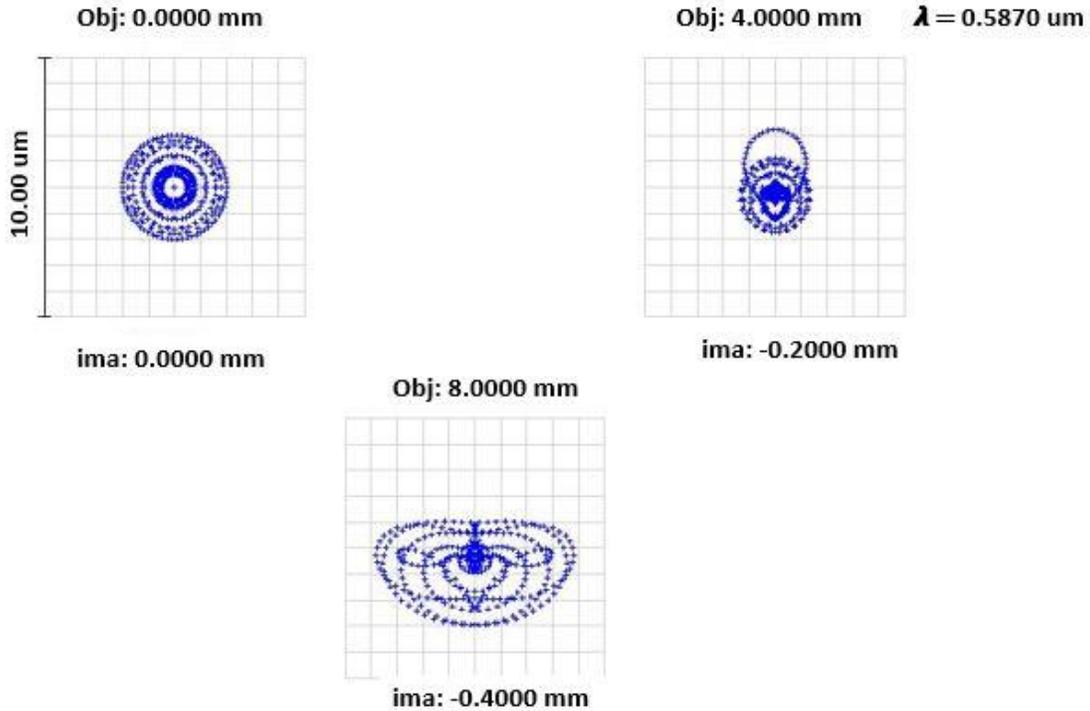

b) Spot diagrams.
Fig. 7 Point at the R2 valley with C=5.3E-4. Lower R2 value.

There is a sensible difference of 2.6 mm in radius R2 between two points having C=5.3E-4, but not any visible difference in spot performance. So, the minimum is not just a point in the parameter space but rather an area with a size that depends on the requirements of the criterion performance.

**V. Search for a deepest point in the R2 valley vicinity.**

The R2 valley minimum that is shown in Table 4 (Fig. 5) was found by traveling over the criterion valley in the direction of R2 radius decrease. To investigate the vicinity of R2 valley minimum for the deepest solution, a new search was conducted. Every eigen vector increment $\Delta C_i$ of the quadratic form of Eq.(5) was analyzed in both directions, the direction of i-th eigen vector and the opposite direction. The step was performed in the direction which gave the deepest decrease of the criterion. After several steps the point with criterion C=5.17E-4, optical prescription of Table 7 and spot diagrams shown in Fig. 8 was found.

**Table 7 Optical prescription at the deepest minimum with C=5.17E-4.**

| Surface | Radius | Thickness | Material | Semi-Diam. |
|---------|--------|-----------|----------|------------|
| OBJ | Infinity | 162.8140 | | |
| STO | Infinity | 0.0000 | | 4.0 |
| 2 | 24.4562 | 2.6182 | K10 | |
| 3 | -8.2221 | 1.0414 | SF1 | |
| 4 | -26.7133 | 0.1524 | | |
| 5 | 11.7542 | 2.1590 | N-SK5 | |
| 6 | Infinity | 7.0579 | | |
| 7 | 4.3285 | 2.5908 | N-SK5 | |
| 8 | -2.9995 | 0.7874 | F5 | |
| 9 | Infinity | 1.3086 | | |
| 10 | Infinity | 0.1778 | N-K5 | |
| 11 | Infinity | 0.0000 | | |
| IMA | Infinity | | | |

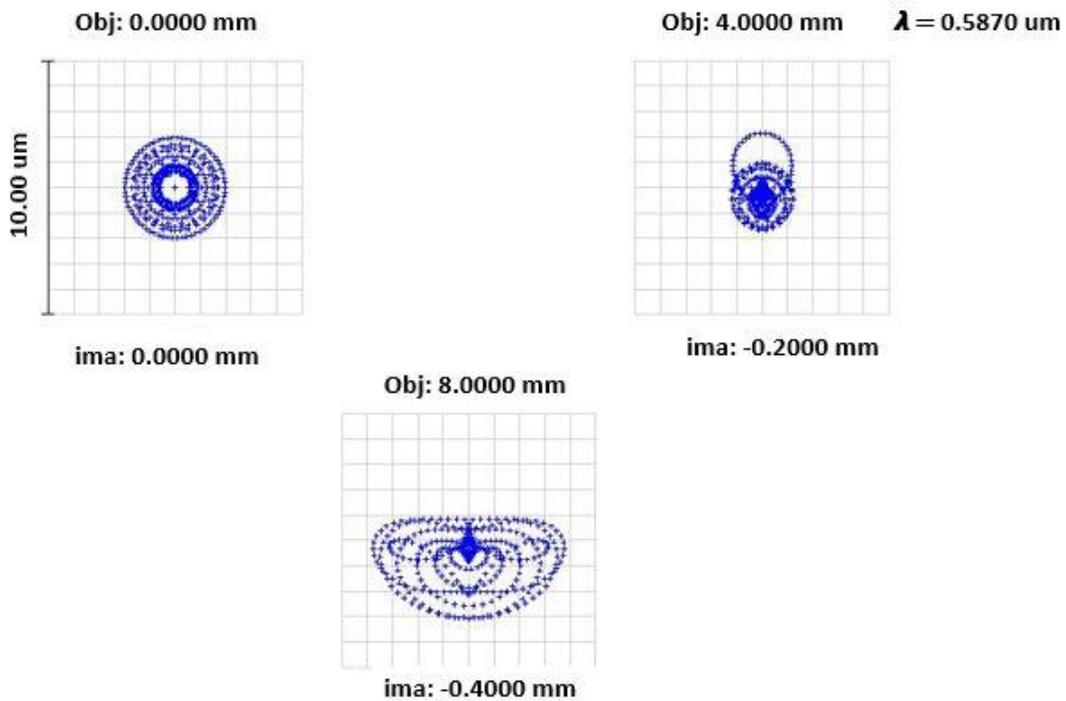

**Fig. 8 Spot diagrams.**

### VI. Travel to the R7= 4 mm R8= -3 mm constraints surface. Search from R7/R8 constraints surface to the opposite side.

The R2 valley search in the direction of decreasing R2 ended at the R3= -7.7 mm, T6=4 mm constraints surface without showing a saddle point. So, further search in this direction is not promising. Another constraints surface closest to the minimum of the R2 valley (Table 4) is R7=4mm and R8= -3 mm surface. Travel over the direction of the R7 reduction at first reveals a minimum with C= 5.17E-4. The absence of saddle point indicates that we are still in the same valley, but just proceeding deeper. This deeper minimum will be conditionally marked as another minimum #2 to simplify perception of the search scheme. The optical prescription of the minimum #2 is shown in Table 8. Moreover, the optical prescription of this minimum and the image quality are very close to the deepest point at the R2 valley shown in the Table 7. This gives some assurance of the reliability of the proposed method.

**Table 8. Optical prescription of the minimum number 2 with C=5.17E-4.**

| Surface | Radius | Thickness | Material | Semi-Diam. |
|---|---|---|---|---|
| OBJ | Infinity | 162.8140 | | |
| STO | Infinity | 0.0000 | | 4.0 |
| 2 | 24.4555 | 2.6182 | K10 | |
| 3 | -8.2139 | 1.0414 | SF1 | |
| 4 | -26.7572 | 0.1524 | | |
| 5 | 11.7407 | 2.1590 | N-SK5 | |
| 6 | Infinity | 7.0390 | | |
| 7 | 4.3406 | 2.5908 | N-SK5 | |
| 8 | -2.9995 | 0.7874 | F5 | |
| 9 | Infinity | 1.3193 | | |
| 10 | Infinity | 0.1778 | N-K5 | |
| 11 | Infinity | 0.0000 | | |
| IMA | Infinity | | | |

Further travel reveals a saddle area with criterion C= 5.28E-4 which is close to landing point at the constraints surface. Then at the transition zone, where penalty function takes it power, all eigen values turn out positive and very close to this landing point the minimum is found with criterion C=5.26E-4. The optical prescription is shown in the Table 9.

Let's step slightly out of the R6 and R7 constraints surface inside the optimization space to neutralize penalty functions influence on eigen vectors. A 3 um (microns) change in R6 and R7 will be enough. Eigen vectors are shown in the Table 10. The first column are increments of criterion along eigen vectors with the step S=0.003. Eigen vector projections on coordinate axis's of prescription parameters are shown as V1-V8. From Table 10, is clear that for the lower increment dC leading out of the constraints surface radius is R8 (V6 are projections of the eigen vector on the axis of surface number 8 curvature). So, stepping out of the constraints surface we are in the R8 "tube" which is guiding us to the positive values of the R8. Traveling over the R8 "tube" passed saddle point with criterion value C= 1.22E-3 and hit the opposite R7= 4mm R8= 3mm constraints surface at the point with C=6.98E-4. The optical prescription at the landing point is shown in the Table 11.

**Table 9. Optical prescription at the minimum on the R7=4mm, R8= -3mm wall. C=5.26E-4.**

| Surface | Radius | Thickness | Material | Semi-Diam. |
|---|---|---|---|---|
| OBJ | Infinity | 162.8140 | | |
| STO | Infinity | 0.0000 | | 4.0 |
| 2 | 22.3243 | 2.6182 | K10 | |
| 3 | -8.5273 | 1.0414 | SF1 | |
| 4 | -27.2037 | 0.1524 | | |
| 5 | 11.7920 | 2.1590 | N-SK5 | |
| 6 | Infinity | 7.4447 | | |
| 7 | 3.9996 | 2.5908 | N-SK5 | |
| 8 | -2.9996 | 0.7874 | F5 | |
| 9 | Infinity | 0.9258 | | |
| 10 | Infinity | 0.1778 | N-K5 | |
| 11 | Infinity | 0.0000 | | |
| IMA | Infinity | | | |

**Table 10. Eigen vectors.**

| | dC | V1 | V2 | V3 | V4 | V5 | V6 | V7 | V8 |
|---|---|---|---|---|---|---|---|---|---|
| 1 | 1.6E00 | 4.6E-1 | 2.4E-1 | -6.6E-1 | 5.4E-1 | 8.8E-2 | 1.1E-3 | 3.8E-3 | 1.4E-2 |
| 2 | 9.0E-4 | -4.2E-1 | 4.1E-2 | 1.3E-1 | 3.6E-1 | 8.2E-1 | 2.4E-1 | -2.3E-2 | 1.9E-2 |
| 3 | 7.6E-4 | -3.8E-1 | 8.9E-1 | 3.0E-2 | -1.4E-3 | -2.4E-1 | 3.0E-3 | 1.4E-4 | -6.6E-3 |
| 4 | -2.5E-5 | 6.5E-1 | 3.8E-1 | 3.5E-1 | -3.6E-1 | 4.2E-1 | 2.6E-2 | -1.7E-2 | 2.8E-2 |
| 5 | 3.2E-5 | 2.1E-1 | -1.1E-2 | 6.5E-1 | 6.6E-1 | -2.8E-1 | 1.3E-1 | 2.0E-2 | 2.5E-2 |
| 6 | 4.6E-6 | -3.8E-2 | -1.4E-2 | -1.0E-1 | -9.1E-2 | 5.2E-3 | 9.7E-1 | 6.7E-2 | 2.0E-1 |
| 7 | 1.3E-6 | 1.7E-2 | 7.4E-3 | -6.0E-4 | 1.3E-3 | 2.9E-2 | 1.9E-1 | 1.8E-1 | -9.7E-1 |
| 8 | 6.0E-7 | 5.0E-3 | -6.4E-3 | -5.1E-3 | 7.7E-3 | -2.7E-2 | 1.0E-1 | -9.8E-1 | -1.6E-1 |

**Table 11. Optical prescription of the landing point at the R7= 4.0 mm R8= 3.0 mm wall.**

| Surface | Radius | Thickness | Material | Semi-Diam. |
|---|---|---|---|---|
| OBJ | Infinity | 162.8140 | | |
| STO | Infinity | 0.0000 | | 4.0 |
| 2 | 19.5445 | 2.6182 | K10 | |
| 3 | -8.5775 | 1.0414 | SF1 | |
| 4 | -25.7804 | 0.1524 | | |
| 5 | 11.8416 | 2.1590 | N-SK5 | |
| 6 | Infinity | 7.3536 | | |
| 7 | 3.9998 | 2.5908 | N-SK5 | |
| 8 | 3.0000 | 0.7874 | F5 | |
| 9 | Infinity | 0.6345 | | |
| 10 | Infinity | 0.1778 | N-K5 | |
| 11 | Infinity | 0.0000 | | |
| IMA | Infinity | | | |

Finally, optimization over the valley with leading variable T6 will find a minimum with C=6.9E-4. The optical prescription is shown in Table 12 and the layout and spot diagrams are shown in Fig. 9.

**Table 12 Optical prescription at the minimum number 3.**

| Surface | Radius | Thickness | Material | Semi-Diam. |
|---|---|---|---|---|
| OBJ | Infinity | 162.8140 | | |
| STO | Infinity | 0.0000 | | 4.0 |
| 2 | 21.6736 | 2.6182 | K10 | |
| 3 | -8.3989 | 1.0414 | SF1 | |
| 4 | -24.1668 | 0.1524 | | |
| 5 | 11.2879 | 2.1590 | N-SK5 | |
| 6 | Infinity | 7.3967 | | |
| 7 | 3.9993 | 2.5908 | N-SK5 | |
| 8 | 3.0000 | 0.7874 | F5 | |
| 9 | Infinity | 0.5524 | | |
| 10 | Infinity | 0.1778 | N-K5 | |
| 11 | Infinity | 0.0000 | | |
| IMA | Infinity | | | |

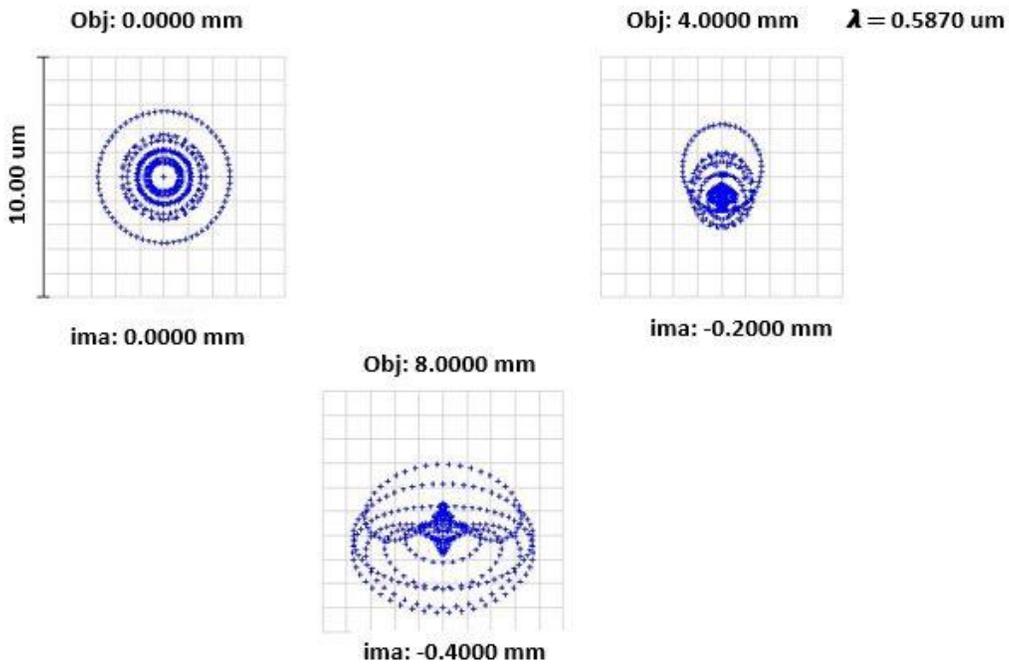

a) Spot diagrams.

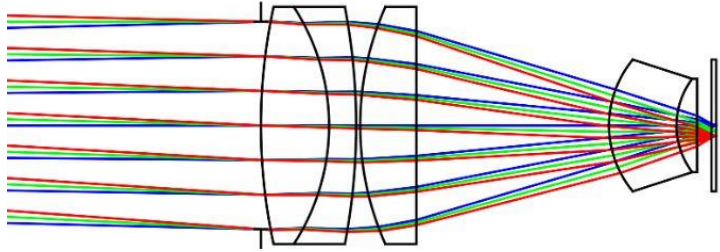

**b) Layout.**
**Fig. 9. Optimal solution at the constrains wall R7= 4.0 mm and R8= 3.0 mm. C=6.9E-4.**

## VII. Navigation from T6=8.2 mm R8= -3 mm constraint surface to a new minimum.

Travel over the main valley in the direction of R2 growth ended at the T6 8.2 mm, R8=-3mm constraints surface. Search for the minimum at the surface opens a new R5 "tube" (valley). Short travel found a new minimum with criterion C= 6.08E-4 shown in the Fig. 10 with optical prescription in the Table 13. A new valley starting from the minimum number 4 is leading to the R7=4.0 mm, R8=3.0 mm constraints surface and then to the minimum number 3 (Fig. 11).

**Table 13. Optical prescription at the minimum number 4**.

| Surface | Radius | Thickness | Material | Semi-Diam. |
|---|---|---|---|---|
| OBJ | Infinity | 162.8140 | | |
| STO | Infinity | 0.0000 | | 4.0 |
| 2 | 61.4631 | 2.6182 | K10 | |
| 3 | -7.7289 | 1.0414 | SF1 | |
| 4 | -19.0863 | 0.1524 | | |
| 5 | 12.1859 | 2.1590 | N-SK5 | |
| 6 | Infinity | 8.1949 | | |
| 7 | 3.9994 | 2.5908 | N-SK5 | |
| 8 | -2.9992 | 0.7874 | F5 | |
| 9 | Infinity | 1.2096 | | |
| 10 | Infinity | 0.1778 | N-K5 | |
| 11 | Infinity | 0.0000 | | |
| IMA | Infinity | | | |

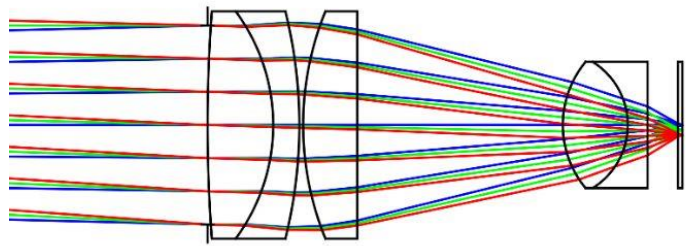

**b) Layout**.

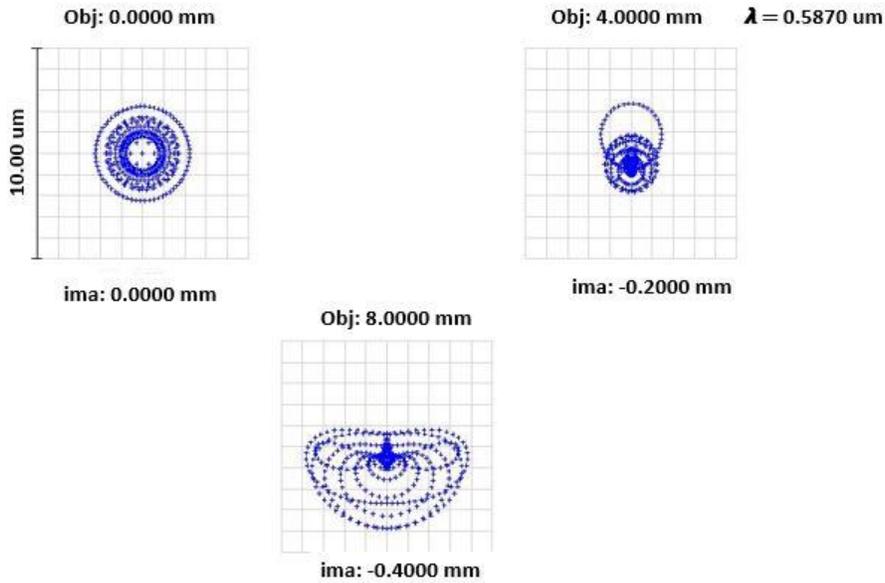

**c) Spot diagrams.**
**Fig 10 Optimal solution on the way out of T6-R8 constraints surface with C=6.08E-4**.

## VIII. Navigation summary.

Here is an illustrative summary of travel through the valleys maze revealing a complete mutually connected system of stationary points (minimums and saddle points).

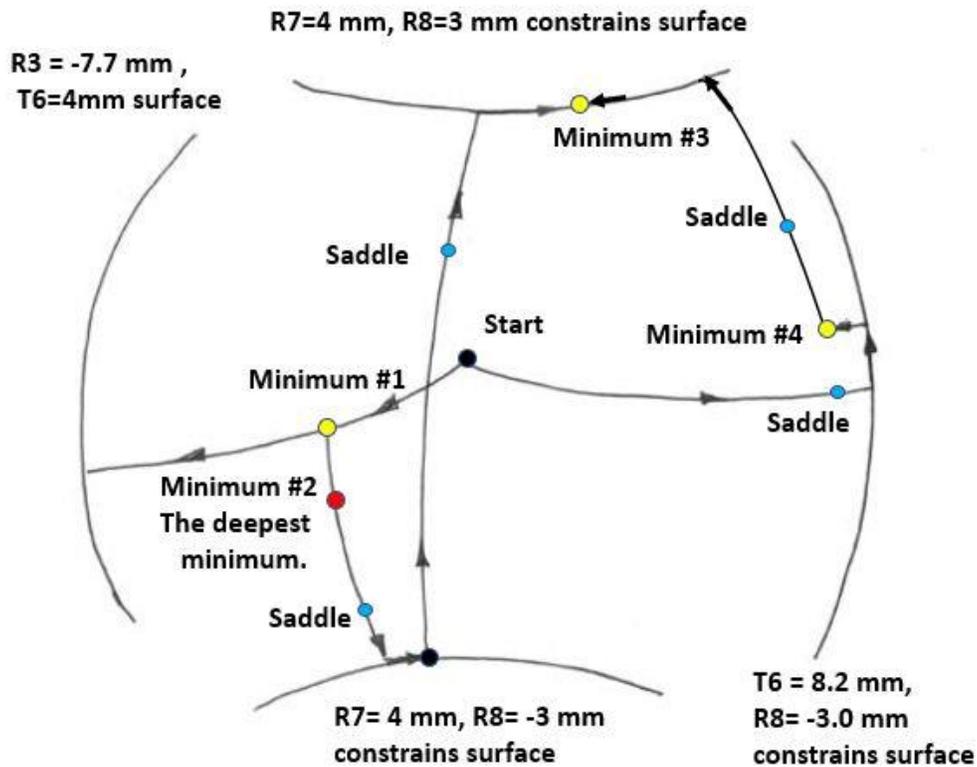

**Fig. 11 Navigation through merit function valleys maze.**

## IX. Optimization with extended waveband

While the development of lasers, laser diodes, and VCSELs has led to an important segment of imaging optics that are designed for narrow wavebands, the majority of practical applications today still require objectives that operate in extended wavebands. Table 14 shows the optical prescription of a redesigned ZEBASE V K002 microscope objective, which was optimized to the nearest merit function valley to operate in an extended waveband of 0.45 - 0.65 µm. Optical glasses in the ZEBASE V prescription were substituted with preferred glasses from the Schott 2019 catalog. The optimization criterion $C_{poly}$ is sum of criterions C of Eq. (2) at three wavelengths: 0.45 µm, 0.587 µm and 0.65 µm. As shown in Table 14, eight optimization parameters are marked as variables with the symbol "v". Constraints on the optimization parameters were shown in the Eq. (1). The layout is shown in Fig.12 and spot diagrams in Fig. 13. The criterion $C_{poly}$ at this point is 3.29E-3.

**Table 14 Optical prescription of the polychromatic version of K002 ZEBASE V objective.**

| Surface | Radius | Thickness | Material | Semi-Diam. |
|---|---|---|---|---|
| OBJ | Infinity | 162.8140 | | |
| STO | Infinity | 0.0000 | | 4.0 |
| 2 | 25.8124 V | 2.6182 | K10 | |
| 3 | -7.9670 V | 1.0414 | SF1 | |
| 4 | -26.2362 V | 0.1524 | | |
| 5 | 11.3779 V | 2.1590 | N-SK5 | |
| 6 | Infinity | 6.7307 V | | |
| 7 | 4.5777 V | 2.5908 | N-SK5 | |
| 8 | -3.1197 V | 0.7874 | F5 | |
| 9 | Infinity | 1.4852 V | | |
| 10 | Infinity | 0.1778 | N-K5 | |
| 11 | Infinity | 0.0000 | | |
| IMA | Infinity | | | |

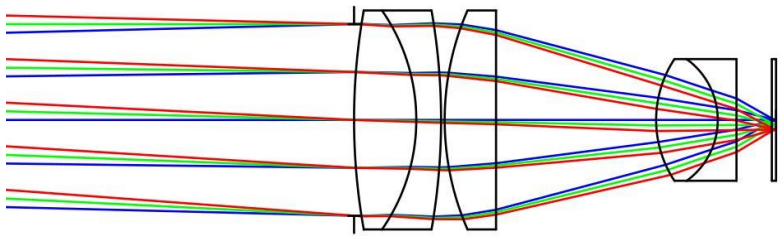

**Fig. 12 Polychromatic version of the microscope objective K_002 from ZEBASEV.**

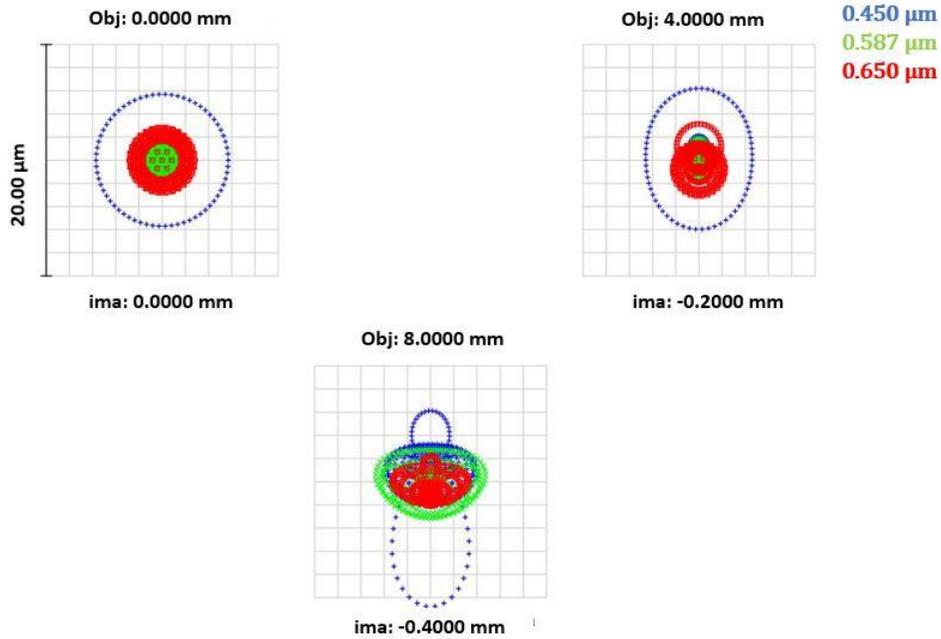

**Fig. 13 Spot diagrams.**

From this point, attempting to optimize over the merit function valley in the direction of decreasing radius R2 soon encounters constraints surface with value of radius R8= -3mm. Further descent over the valley within this constraint led to an optimal point with criterion $C_{poly}$= 2.84E-3. The optical prescription for this optimal point is shown in Table 15, the layout is shown in Fig. 14, and spot diagrams are shown in Fig. 15.

**Table 15 Optical prescription of the optimal point with $C_{poly}$= 2.84E-3.**

| Surface | Radius | Thickness | Material | Semi-Diam. |
| --- | --- | --- | --- | --- |
| OBJ | Infinity | 162.8140 | | |
| STO | Infinity | 0.0000 | | 4.0 |
| 2 | 20.9872 | 2.6182 | K10 | |
| 3 | -8.4019 | 1.0414 | SF1 | |
| 4 | -31.1107 | 0.1524 | | |
| 5 | 11.6116 | 2.1590 | N-SK5 | |
| 6 | Infinity | 6.8475 | | |
| 7 | 4.3968 | 2.5908 | N-SK5 | |
| 8 | -3.0000 | 0.7874 | F5 | |
| 9 | Infinity | 1.3339 | | |
| 10 | Infinity | 0.1778 | N-K5 | |
| 11 | Infinity | 0.0000 | | |
| IMA | Infinity | | | |

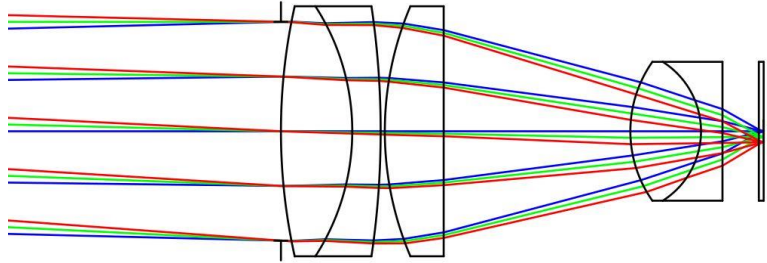

**Fig. 14 Layout at the optimal point.**

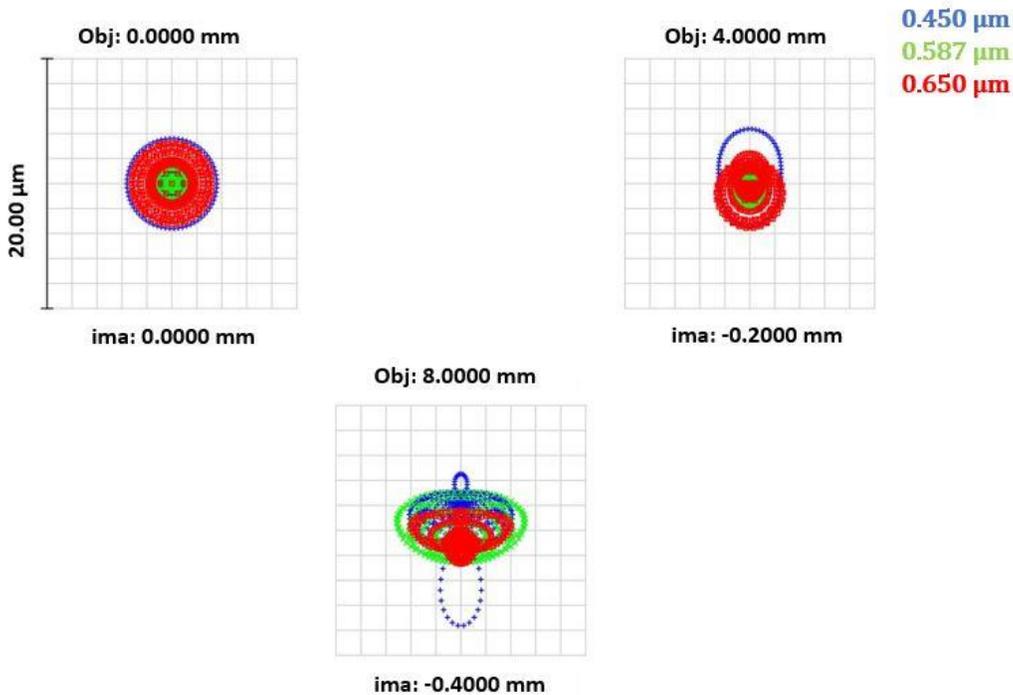

**Fig. 15 Spot diagrams.**

Next, consider optimization over the valley in the direction of increasing radius R2. This first leads to the constraint surface R8= -3mm, followed by constraint surface T6= 8.2 mm, at which point the criterion value $C_{poly}$= 3.75E-3. Further descent over the valley within the six-dimensional constraint surface (R8, T6) leads to an optimal point with criterion $C_{poly}$= 3.71E-3. The optical prescription for this point is shown in Table 16, the layout is shown in Fig. 16, and the spot diagrams are shown in Fig. 17.

If we will move in the direction of reducing T6 from the point with prescription of the Table 16 (keeping R8 = -3mm), we would be led back to the minimum of the Table 15. This means we have two valleys connecting these two points, where each valley does not have a saddle point. Thus, the minimum at the constraint surface (R8, T6) is not an independent minimum. Such minimums are formed at the constraint surface when it crosses a full-size valley in the unconstrained space. Another valley from the optimal point with $C_{poly}$= 2.84E-3 (Table 15) in the direction of increasing radius R8 leads through a saddle point to the minimum with prescription similar to the Table 12. However, this minimum has poor quality. The criterion at this point is $C_{poly}$= 1.41E-2.

**Table 16** Optical prescription of the optimal point with $C_{poly}$= 3.71E-3 at the constraints surface (R8, T6).

| Surface | Radius | Thickness | Material | Semi-Diam. |
|---------|--------|-----------|----------|------------|
| OBJ | Infinity | 162.8140 | | |
| STO | Infinity | 0.0000 | | 4.0 |
| 2 | 51.9012 | 2.6182 | K10 | |
| 3 | -7.9053 | 1.0414 | SF1 | |
| 4 | -21.3046 | 0.1524 | | |
| 5 | 12.3816 | 2.1590 | N-SK5 | |
| 6 | Infinity | 8.2000 | | |
| 7 | 4.0651 | 2.5908 | N-SK5 | |
| 8 | -3.0000 | 0.7874 | F5 | |
| 9 | Infinity | 1.4145 | | |
| 10 | Infinity | 0.1778 | N-K5 | |
| 11 | Infinity | 0.0000 | | |
| IMA | Infinity | | | |

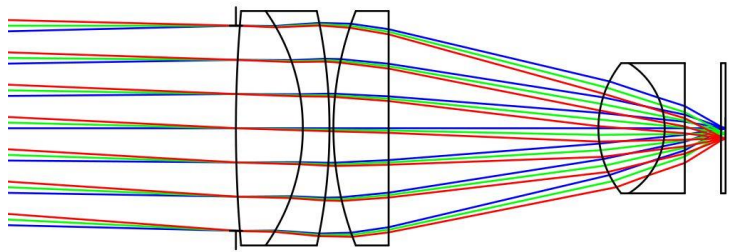

**Fig. 16** Layout of optimal point at the (R8, T6) constraints surface.

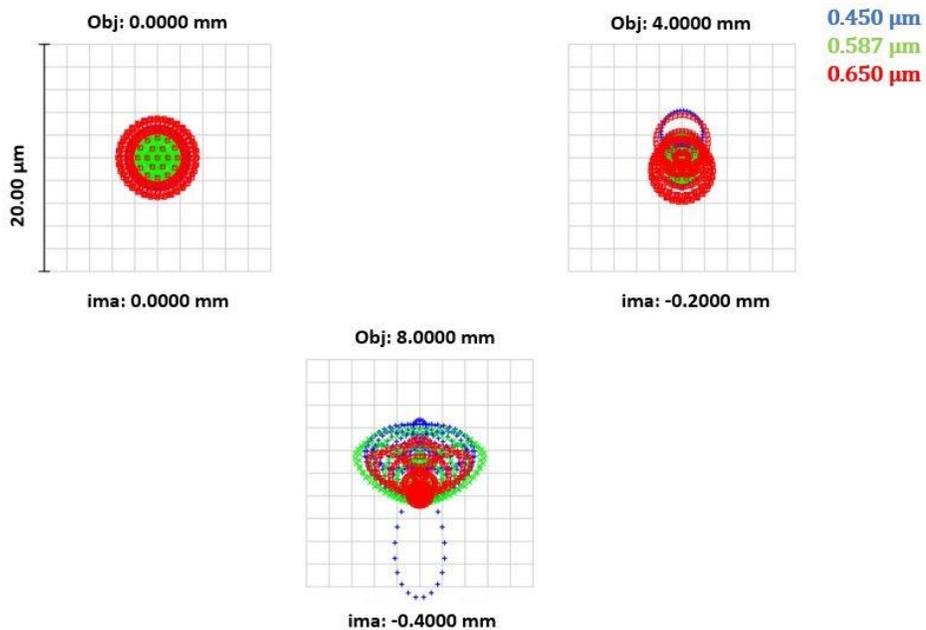

**Fig. 17** Spot diagrams

# VIII. Conclusion.

In this paper was shown that local minimums in optimization of even moderate complexity optical systems are located over merit function valleys and specific algorithms for traveling over such valleys were proposed. No sensible difference in the criterion values over several mm's in radii or air gaps at valleys bottom was found. So these minimums can not be considered as some points in the multidimensional optimization space but are areas at the valley bottoms. Sometimes these valleys are so long that the choice of a solution can be based on manufacturability/cost criterions. These extended valley areas of solutions do not relax tolerances because at every point a fine combination of prescription parameters is required. A special role of saddle points as a point of separation of attraction areas to neighboring minimums was clarified. It was shown that the gradient curve connecting neighboring minimums through the saddle point has a lowest maximum value of optimization criterion besides other such curves connecting these minimums. So, these gradient curves are the path of the slowest growth/descent and therefore are the merit function valleys. An efficient algorithm to travel over these valleys from the one minimum to a saddle point and further to a new minimum was proposed.

Practical optimization in lens design is associated with number of constraints on parameters. In the constrained optimization space, valleys of the criterion function can avoid mutual intersections. Connections between valleys can be found on the constraint surfaces. Local minimums at the constraint surfaces will be valley footprints. Each new minimum will be an entrance to the new valleys/tunnels leading to other criterion minimums. Results of multi-extremum optimization of the microscope objective demonstrated an efficiency of the proposed algorithms. Extended work with optimization of other type of lens architectures [15] has to be preformed to mature the algorithm. The relationship between Seidel aberration theory [16] and multi-extremum optimization results has to be clarified as well.

Optimization algorithms required calculations of the first and second derivatives of the criterion in regard to optimization parameters. In this research, raytracing tests of derivatives were used. However, in the future analytical derivatives tests [17, 18] can improve the accuracy and accelerate proposed optimization procedures. The proposed optimization algorithms in this paper, are associated with extended computer burden and cannot be recommended to lens designers for everyday use. But further progress in computer CPU's clock speed together with implementation of multiple core parallel processing will probably make it possible to use this method of global optimization in commercial lens design software soon enough.